# Solute Induced Jittery Motion of Self-Propelled Droplets

Prateek Dwivedi[a], Bishwa Ranjan Si[a], Dipin Pillai[a] and Rahul Mangal[a*]

[a] Department of Chemical Engineering, Indian Institute of Technology Kanpur, Kanpur-208016, India

**ABSTRACT:** The intriguing role of the presence of solutes in the activity of a self-propelling droplet is investigated. A system of self-propelling micron sized 4′-pentyl-4-biphenylcarbonitrile (5CB) droplets in an aqueous solution of tetradecyltrimethylammonium bromide (TTAB) as surfactant is considered. It is shown that addition of glycerol causes the active 5CB droplet to exhibit a transition from smooth to jittery motion. The motion is found to be *independent* of the droplet size and the nematic state of 5CB. Analogous experiments with Polyacrylamide (PAAm), Polyvinylpyrrolidone (PVP) and Polyvinyl Alcohol (PVA), as solutes confirm that such a transition cannot merely be explained solely based on the viscosity or Peclet number of the system. We propose that the specific nature of physicochemical interactions between the solute and the droplet interface is at the root of this transition. The experiments show that the time-scales associated with the influx and redistribution of surfactants at the interface are altered in the presence of solutes. Glycerol and PVP significantly enhance the rate of solubilization of the 5CB droplets resulting in a quicker re-distribution of the adsorbed TTAB molecules on the interface, causing the droplet to momentarily stop and then restart in an independent direction. On the other hand, low solubilization rates in the presence of PAAm and PVA lead to smooth trajectories. Our hypothesis is supported by the time evolution of droplet size and interfacial velocity measurements in the presence and absence of solute. Overall, our results provide fundamental insights into the complex interactions emerging due to the presence of solutes.

INTRODUCTION

Micro-organisms, characterized by very small length-scales, exhibit thermal fluctuation induced Brownian motion, which however cannot result in any form of directional propulsion. Furthermore, at these length scales, viscous forces typically dominate inertia (Reynolds number, Re<<1) and their motion is governed by the *reversible* Stokes equation. Therefore, certain unique and non-reciprocal strategies, different from the mechanisms used by macro-organisms, are required for microbial propulsion. Micro-organisms such as bacteria and unicellular protozoa utilize appendages attached to their molecular motors, which are driven by the chemical energy they extract from the surrounding medium, to generate a net directional propulsion.[1,2] This unique ability of micro-organisms to "self-propel" themselves has intrigued the scientific community over decades and in an attempt to fundamentally understand their motion, numerous theoretical and computational studies have been undertaken.[3–9] Lately, researchers have also started exploring artificial model systems known as "artificial active matter" or "artificial micro-swimmers", which mimic the life-like motion of micro-organisms.[10–12] Unlike equilibrium Brownian motion, these artificial active systems exhibit unique transport characteristics. Upon carefully tailoring their design, these artificial systems can transport cargoes,[13–16] perform intricate tasks in microscopic domains[17–19] and find applications in tuning the bulk material properties.[20,21] Therefore, apart from providing fundamental insights into the out-of-equilibrium motion, these model systems continue to attract tremendous interest due to their potential applications in diverse fields such as medicine, material and environmental science. [22–26]

These artificial active matter systems do not rely on any externally imposed force fields for their propulsion. Instead, they use a local (particle level) gradient generated by their asymmetric interactions with the surrounding medium. One way, to invoke the desired asymmetry is via incorporating an intrinsic asymmetry (shape or composition) in the particle itself. Janus colloids, in which two half sides of the surface of the particles are of different chemical compositions, are widely used for self-propulsion. Their chemical inhomogeneity generates an asymmetric interaction with the surrounding medium, which drives the particle by a mechanism known as self-diffusiophoresis.[12,27,28] Another such class of artificial active matter is that of "active emulsions"[10,29–32], where droplets of one fluid perform self-propelled motion while dispersed in another immiscible fluid. The most explored system is that of oil/water droplets in surfactant filled water/oil medium. In these active emulsions, symmetry is 'spontaneously' broken in an otherwise isotropic spherical droplet, by generating a non-uniformity in the surfactant concentration along the interface via some mechanism. Asymmetric coverage of the surfactant at the interface creates a gradient in the interfacial tension ($\gamma$), which drives the fluid from low $\gamma$ toward high $\gamma$, resulting in the well-known Marangoni flow. Since, there is no net external force acting on the system, to conserve the overall linear momentum, this interfacial flow propels the droplet toward the region of low interfacial tension. In contrast to solid active Janus particles, these active droplets can deform[33,34] while performing robust active motion. Therefore, these systems are considered as excellent model systems to understand microbiological swimmers. In addition, due to their ability to exhibit chemotaxis[30] and their bio-compatibility, these droplets are compelling candidates for active matter engineering including drug delivery and other bio-medical applications.[26]

Over the years, different approaches to invoke spontaneous asymmetry in isotropic droplets with uniform surfactant coverage have been reported. One approach is to use localized targeted chemical reactions such as hydrolysis[35,36] and bromination[31,37] of surfactant molecules, which alter their surface activity. Numerous experimental reports have utilized this approach of interfacial chemical reaction to achieve self-propulsion of droplets.[32,38,39] Another common strategy is to utilize the adsorption-depletion of surfactants triggered by micellar solubilization.[30,40–42] Different combinations of oil droplets in aqueous ionic surfactant solutions[10,29–36,43] and inverse oil in water emulsions[17] have been reported to display self-propelled motion through this mechanism. Employing a nematic liquid crystal (LC) phase as the active oil droplet is known to help in visualization of the flow-field inside the droplets. Experimentally, 4-pentyl-4'-cyanobiphenyl (5CB), a thermotropic liquid crystal, in aqueous solution of ionic tetradecyltrimethylammonium bromide (TTAB) surfactant has been widely studied.[30,44–46] Using 5CB in TTAB aqueous solution as the model system, past studies have investigated a few fundamental aspects of active droplet motion, including the effects of the surfactant concentration and size of the droplet.[44] Kruger et al.[46] demonstrated that the nematic phase of the droplet causes it to adopt a novel curling mode of motion.[46] These droplets have also been demonstrated to exhibit chemotaxis and negative auto-chemotaxis,[30] rendering them the most ideal system to model biological microorganisms and cellular motion.

The presence of a non-reactive molecular solute in the continuous phase is expected to significantly affect the activity of the self-propelling droplet. Addition of such a solute is expected to alter various properties of the system including viscosity, pH, interfacial tension, rate of droplet solubilization among others, which can have a profound effect on the self-propelled motion. However, this aspect continues to remain relatively unexplored. Thus, given the underlying complex behavior and rich physics, further experimental studies are warranted towards systematically characterizing these mode transitions and the underlying reasons thereof. Furthermore, to realize the goal of using these artificial active droplets in performing intricate tasks, it is critical to understand the effect of these additional solutes. To the best of our knowledge, there is only one very recent arxiv report by Hokmabad *et. al.*[47] , which experimentally reported an onset of jittery motion in 5CB droplets in presence of glycerol as a solute. The observed jittery motion was attributed to an increase in the ambient fluid viscosity and thereby Peclet number (*Pe*). Based on linear stability analysis, they showed that a symmetry-breaking translational instability occurs at *Pe*~4. A further increase in Peclet number (*Pe*>90) leads to a dominance of higher order spherical harmonic modes, resulting in a quad-polar flow profile, that induced jittery motion. In this study, we experimentally demonstrate that the role of solute is indeed more complex than hitherto understood. Using separate experiments with glycerol, Polyacrylamide (PAAm), Polyvinylpyrrolidone (PVP) and Polyvinyl Alcohol (PVA), as solutes, we show that these systems at similar *Pe* exhibit different self-propelling activity. Under similar *Pe* conditions, while jittery motion is induced in the presence of glycerol and PVP, such a transition is never observed in the presence of PAAm and PVA. The difference in the droplet activity induced by different solutes is explained based on the role of solute in altering the rate of micellar solubilization which affects the competing times scales of influx and outflux of surfactant molecules on the interface. We bring forth the intriguing physics behind the jittery motion induced by the addition of solutes in the active motion of 5CB oil droplets.

EXPERIMENTAL SECTION

We dispersed 4- pentyl-4′-cyanobiphenyl (5CB, Sigma Aldrich) droplets in aqueous solution of 6wt% trimethyl ammonium bromide (TTAB, Loba chemicals). Glycerol (Loba Chemicals), Polyacrylamide (PAAm) of molecular weight ($M_w$)~6000 kDa, (Polysciences Inc.), Polyvinylpyrrolidone (PVP) of $M_w$~40 kDa (Loba Chemicals) and Polyvinyl Alcohol (PVA) of $M_w$~115 kDa (Loba Chemicals), were used as molecular solutes in separate experiments. 5CB droplets of size ~20μm-100μm were produced by vortex mixing the solution. Low concentration of 5CB, ~ 0.15 v/v%, was used to ensure low number density of droplets to avoid droplet-droplet interaction.

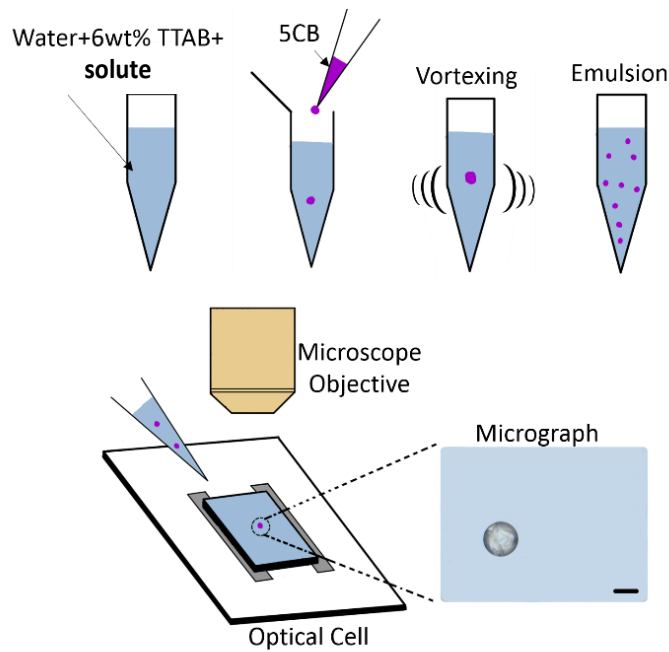

Figure 1: Schematic representation of the experimental setup. The scale bar in the optical micrograph corresponds to 40μm.

The emulsion was injected into a custom-made optical cell with vertical height of 100 μm. Motion of isolated 5CB droplets was observed using upright polarized optical microscope Olympus BX53. An Olympus LC30 camera with 2048 x 1532 pixels resolution was used to record the motion of the droplets. A thermal stage was mounted on the microscope stage to maintain a constant temperature during the measurements. Droplet tracking was performed with Image-J software, using an image correlation-based approach, to obtain particle trajectories (X (μm), Y (μm)) vs. elapsed time ($\Delta t$). Figure 1 is a schematic representation of the experimental setup. Interfacial tension of 5CB in aqueous solution with varying TTAB concentration ($c_{TTAB}$) and on addition of different solutes was measured at 25 °C using pendant drop contact angle meter (Data Physics OCA 35). The interface shape of the pendent LC droplet in the aqueous solution was fitted using Young-Laplace equation. A rheometer (TA Instruments DHR3) with a Couette assembly was used to measure the viscosity of the aqueous solutions with solutes.

RESULTS AND DISCUSSION

Mechniasm of self-propulsion

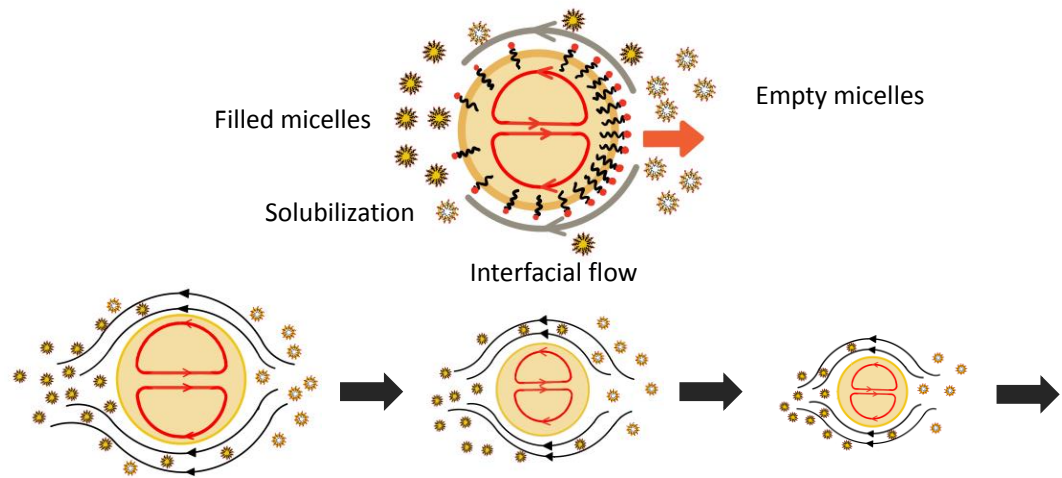

Figure 2: Schematic illustration of the micellar solubilization induced self-propulsion of an oil droplet in aqueous surfactant solution.

Using the schematic shown in figure 2, we first briefly review the mechanism of micellar solubilization and consequent self-propulsion of the droplets.[17,29,34,45,48–50] Solubilization of a stationary oil droplet in an aqueous surfactant solution may occur either due to molecular dissolution of oil into the ambient medium or due to micellar solubilization. For ionic surfactants, molecular dissolution at the interface is negligible due to electrostatic repulsion between adsorbed surfactants on the interface and approaching empty micelles. In such systems, the physics is therefore governed predominantly by micellar solubilization. Based on a theoretical analysis, Todorov *et al.*[50] concluded that a constant rate of solubilization, independent of droplet size, is indicative of rate-limited micellar solubilization within a narrow solubilization boundary layer. In our experiments, we observe a constant rate of solubilization (discussed later) of 5CB droplets in different dispersing media, independent of the droplet size, which indicates that our system also exhibits reaction-limited micellar solubilization. Further, as the droplet undergoing micellar solubilization exhibits self-propulsion in our experiments, droplet convection begins to play a dominant role. The spontaneous breakup of symmetry can only be sustained by an additional physics of advective influx/outflux of surfactants/micelles, which may be understood simplistically as a sequence of following events. (i) Since the surfactant concentration is significantly greater than CMC, the droplet easily encounters a free micelle from any arbitrary direction. The micelle dumps surfactants locally at the interfacial region reducing the local interfacial tension ($\gamma$). By incorporating some of the surfactant molecules in the interfacial region, a tiny oil filled micelle, leaves the parent droplet. This process of *micellar solubilization* disturbs the otherwise symmetric surfactant coverage and generates an instantenous interface surfactant concentration gradient, i.e., $\nabla_s c$. (ii) This inhomogeneous surfactant coverage generates Marangoni stress along the interface which drives the fluid carrying the surfactant from surfactant-rich region to surfactant-lean region. The interfacial flow is accompanied by the surrounding fluid flow in the same direction. To conserve the overall linear momentum, the droplet self-propels in the opposite direction. (iii) Once the droplet starts moving, it encounters more free micelles at the leading edge compared to the trailing edge. The additional surfactant molecules acquired by the collisions with free micelles get transferred towards the trailing edge due to the Maragoni effect. Therefore, after the onset of the active motion, solubilization mainly takes place from the posterior region of the droplet. (iv) The higher solubilization in the trailing region further lowers the surfactant concentration. At the same time, the leading edge gets fresh supply of surfactants by encountering free micelles via advection. Therefore, a gradient ($\nabla_s c$) is maintained which sustains the active motion. Theoretically, it has been shown that above a critical Peclet number, a stationary isotropic Brownian droplet can become linearly unstable, leading to an autonomous sustained active motion.[41,42]

Motion in TTAB-water mixture

We first benchmark our experimental setup by investigating the motion of 5CB liquid crystal droplets in $c_{TTAB}$ =6wt% at a constant temperature of 25°C. Since, it is known that micelles play a key role in the propulsion mechanism, maintaining the bulk surfactant concentration greater than CMC is a necessary condition for active propulsion.[10,29,30] The CMC for TTAB in water is ~ 0.13wt%, and therefore, our experiments with $c_{TTAB}$ =6wt% ensures that the bulk TTAB concentration is maintained much higher than the CMC. As the height of the optical cell is of the same order of the droplet size, their motion remains mostly constrained in the 2D (X-Y) plane with occasional drift in z direction. In figure 3(a), we show a few representative X-Y trajectories for different sizes of active 5CB droplets observed for 100 seconds. Absence of any directional bias in the trajectories confirms negligible bulk convection. The trajectories also indicate a size-dependent transition in the mode of active motion of the droplets, with larger droplets (50μm-70μm) performing random active motion, moderate size (20 μm-50 μm) droplets tracing out spiral trajectories (curling motion) and smaller droplets (<20 μm) displaying relatively linear motion. These observations are also supported by their corresponding temporal evolution of the 2D mean square displacement (MSD), shown in figure 2(b). MSD values were obtained by tracking the X-Y position of the droplets' center of mass at different time intervals ($\Delta t$). For a small droplet (~19 μm), which performs linear motion, the MSD scales as $\Delta t^2$ at all time scales. For a medium sized droplet (~33 μm) performing curling motion, the MSD scales as $\Delta t^2$ at both short and very long timescales. However, at intermediate time scales, a saturation in MSD is observed, which is characteristic of the curling motion, which has been attributed to the viscous forces deforming the director field of the nematic 5CB droplets. For the largest droplet (~70 μm) which performs random active motion, the MSD scales as $\Delta t^2$ at short time scales, but transitions to a linear scaling of $\Delta t^1$ at long times. Increasing the droplet size results in a higher magnitude of MSD for same $\Delta t$, suggesting higher propulsion speed for larger sized droplets. These foregoing observations of the control experiments are in good agreement with previous studies.[44,46]

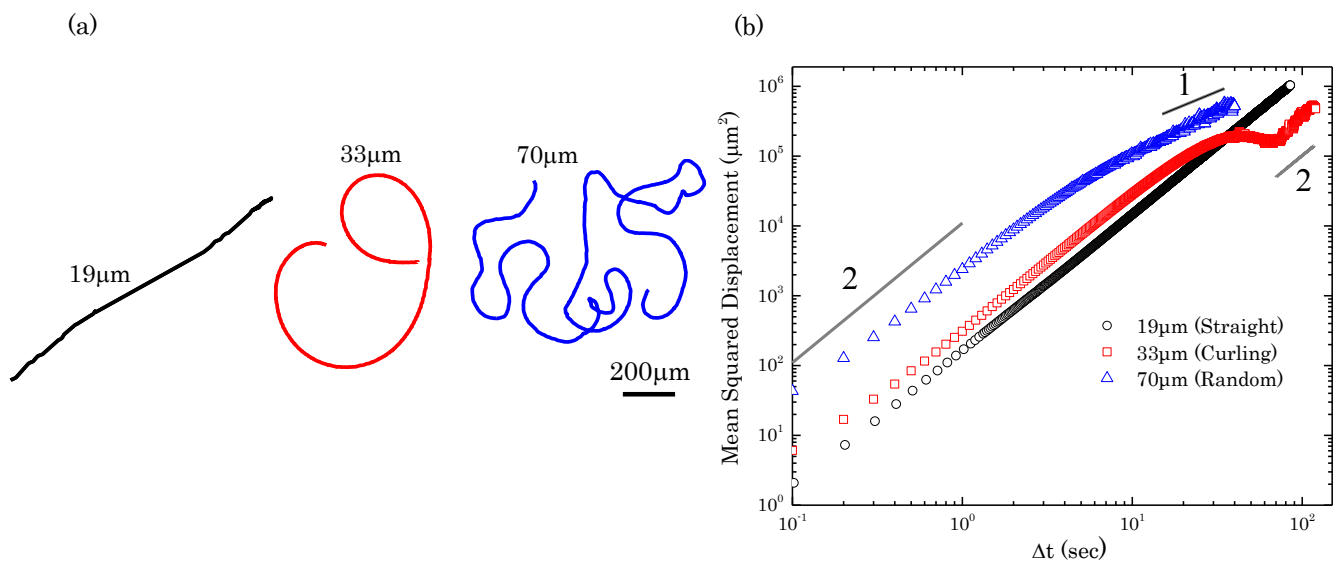

Figure 3: 5CB droplets in 6wt% TTAB aqueous solution: (a) Trajectories ($\Delta t$=100 s) of active droplets with varying size. 19 μm droplet shows linear, 33 μm droplet shows curling and 70μm droplet shows random trajectory. (b) Mean square displacement with time for trajectories captured over 1000s.

Effect of solutes

To investigate the effect of addition of a molecular solute on the motion of active 5CB droplets, we first gradually added varying concentrations ($x_{Gly}$=20wt%, 50wt%, 60wt% and 80wt%) of glycerol to the surfactant-laden aqueous solution. The CMC for $x_{Gly}$=80wt% solution was determined by measuring the interfacial tension with TTAB concentration (see supporting figure S1). The $CMC_{80wt\%Glycerol-water}$ is 0.45wt%, which is greater than $CMC_{water}$, viz., 0.13 wt%, which is consistent with previous reports.[51] A bulk TTAB concentration of 6wt%, therefore, ensures the $c_{TTAB}$ >> CMC for all glycerol concentrations, and, hence, satisfies the necessary condition for self-propulsion. Figure 4(a), shows a few representative X-Y trajectories for active 5CB droplet (~50μm) in a glycerol-TTAB-water solution with varying glycerol concentration (see supporting videos S1-S3). It is seen that addition of 20wt% glycerol does not affect the motion significantly. However, at higher concentrations, onset of unusual jittery fluctuations in an otherwise smooth active trajectory is evident. Representative videos are shown in supporting movies S4-S6. Similar observations on the addition of glycerol have been concurrently reported by Hokmabad *et al.*[47]

The Brownian translational fluctuation ($\Delta L \sim \sqrt{4D\Delta t}$) for a similar-sized droplet for $\Delta t$=100s comes to be around ~0.16-0.28μm, which is significantly smaller than the observed fluctuation length of ~10-50 μm. Here, $D = \frac{k_B T}{6\pi\eta R}$, is the Brownian Diffusivity obtained using classical Stokes-Einstein equation, $k_B$ is the Boltzman constant, $T$ is the absolute temperature, $\eta$ is the dynamic viscosity of the continuous phase and $R$ is the droplet radius. For active colloids, the change in the direction of motion is influenced by the Brownian rotational diffusivity $D_r = \frac{kT}{8\pi\eta R^3}$.[27] Since, addition of glycerol results in an increase in viscosity of the aqueous phase, the onset of this peculiar jittery motion, is in contrast to the theoretical expectation of reduced rotation in a more viscous media. Hence, it is evident that the observed fluctuations are non-Brownian in nature. In active droplets, interfacial shape deformation may also contribute to a change in direction of active transport. However, an increase in viscosity of the continuous phase due to glycerol is only expected to suppress any such deformation, suggesting that that the observed fluctuations do not owe their origin to interfacial instability. We also carried out experiments at T=42°C, which is above the nematic to isotropic transition temperature (35°C) of 5CB.[52] The motion of the isotropic 5CB droplet is similar to the fluctuating motion observed at 25 °C, see supporting figure S2 and movie S7. Based on these observations, we conclude that the nematic phase of the 5CB droplet, which earlier has been shown to play a critical role in governing the mode of active transport in aqueous surfactant solution,[44,46] is not responsible in inducing the jittery motion. We also observed that in pure water, the 5CB droplets perform motion close to the bottom wall of the optical cell, whereas, in the presence of glycerol they remain close to the top surface. This behavior of 5CB droplets is consistent with the density values $\rho_{water}$ ~ 0.99 g ml$^{-1}$, $\rho_{LC}$ ~ 1.057 g ml$^{-1}$ and $\rho_{glycerol}$ ~ 1.25 g ml$^{-1}$ at 25°C. We also repeated the experiment in heavy water D$_2$O ($\rho_{D_2O}$ ~ 1.11 g ml$^{-1}$), wherein, similar to aqueous glycerol solution, LC droplets are lighter and remain close to the top wall. Absence of jittery motion in D$_2$O (supporting video S8) further confirms that the observed jittery motion in glycerol is not due to the interaction with wall and the confinement. We believe that the underlying mechanism is solute induced, which we shall expand upon later.

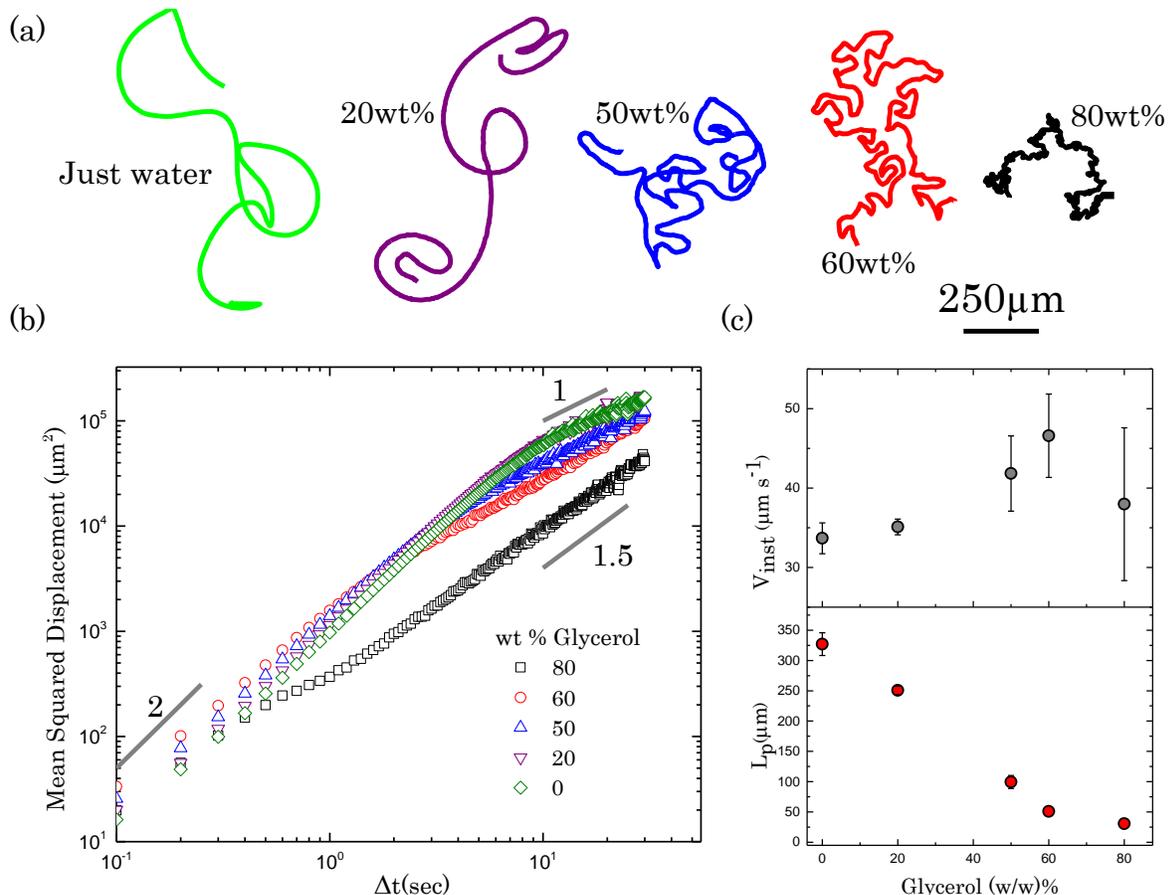

Figure 4: (a) Trajectories ($\Delta t$=100 s) of active 5CB droplet (~50 μm) in glycerol-water TTAB solution with varying glycerol wt%. (b) Mean square displacement with time for varying glycerol wt% for trajectories captured over 600s. (c) Instantaneous speed and persistence length of 5CB droplets with varying glycerol concentration.

From figure 4(a), it can also be observed that the jitteriness in the trajectory increases continuously with increase in $x_{Gly}$. To quantify the intensity of fluctuations, we compute the persistence length of the trajectories given by $L_p = \langle v_{inst.} \rangle \tau_R$. Here, $\langle v_{inst.} \rangle = \left\langle \left| \frac{\mathbf{r}_{i+1} - \mathbf{r}_i}{t_{i+1} - t_i} \right| \right\rangle$ is the mean instantaneous speed of the droplet calculated using the time dependent position ($\mathbf{r}_i(t_i)$) and $\tau_R$ is the inverse of the rotational diffusivity ($D_r$) of the droplet. $\tau_R$ is obtained by fitting the velocity autocorrelation $C(\Delta t) = \langle \mathbf{v}_{inst.}(\Delta t) \cdot \mathbf{v}_{inst.}(0) \rangle$ using the following equation[53,54] (see supporting figure S3-S4):

$$C(\Delta t) = 4D\delta(\Delta t) + \langle v_{inst.} \rangle^2 \cos(\omega \Delta t) \exp(-\Delta t / \tau_R) \ldots\ldots\ldots\ldots(1)$$

Here, $\mathbf{v}_{inst.} = \frac{\mathbf{r}_{i+1} - \mathbf{r}_i}{t_{i+1} - t_i}$ is the instantaneous velocity, $D$ is the Stokes-Einstein diffusivity of the droplet, $\delta$ is the Dirac-delta function and $\omega$ is the angular velocity. A monotonic decrease in persistence length ($L_p$) with glycerol concentration, as shown in figure 4(c), supports the observation in figure 4(a) that the trajectories become more jittery at higher $x_{Gly}$. The corresponding 2D MSD (X,Y) vs $\Delta t$ shown in figure 4(b) indicates a scaling of $\Delta t^2$ at short timescales for the different glycerol concentrations, which is consistent with the active motion of the droplets. Similarity of the droplet active motion at low $x_{Gly}$ with the motion in pure water, is reflected as MSD ~ $\Delta t^1$ at long times, implying random motion. With increasing $x_{Gly}$, at long timescales, MSD vs $\Delta t$ exhibits a gradual transition in scaling from $\Delta t^1$ to $\Delta t^{1.5}$. This shift in the scaling of MSD suggests that at higher $x_{Gly}$ active 5CB droplets perform a *self-avoiding random walk*.

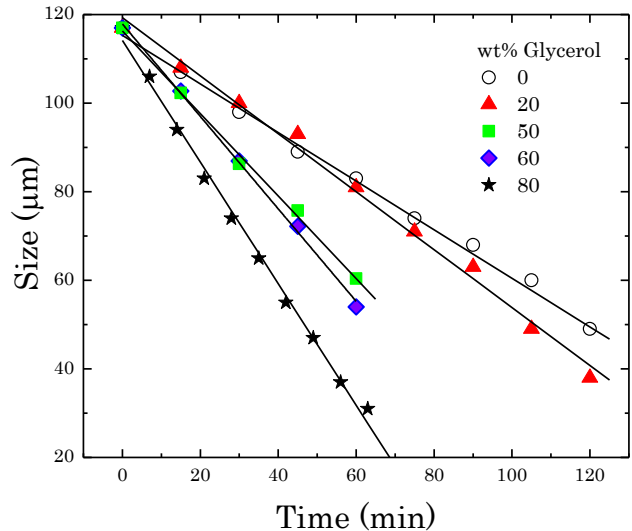

Figure 5: Temporal variation of the size of active 5CB droplets with varying glycerol concentrations in the aqueous phase.

To further understand these observations, we calculate the rate of solubilization (-dR/dt) of the droplets with varying $x_{Gly}$, see figure 5. Continuous shrinkage of the active 5B droplet has been attributed to micellar solubilization, a phenomenon essential for active motion. For all $x_{Gly}$, the droplet shrinks linearly with time in the course of its self-propelled motion, which is consistent with existing reports.[45] From the figure, it is also evident that increasing $x_{Gly}$ results in a faster solubilization of the active 5CB droplets, suggesting that with more glycerol an actively moving 5CB droplet will release a higher number density of filled micelles in its trail.

Jin *et al.*[30] earlier demonstrated that an active 5CB droplet avoids its own trail of filled micelles exhibiting negative autochemotaxis. In solute-free aqueous TTAB solution, since the persistence length, $L_p$, is large, the long-time motion is purely random with the MSD scaling as $\Delta t^1$. This is because at such long time scales, the filled micelles disperse away allowing the droplet to visit a previsited location. This is because at such long time scales, the filled micelles disperse away allowing the droplet to occupy a previsited position. With an increase in glycerol concentration, the rate of solubilization increases due to which the 5CB droplets leave a denser trail of filled micelles. Also, $L_p$ is smaller at higher glycerol concentrations, forcing them to change directions more frequently. These factors explain the observation that at higher glycerol concentrations, a droplet will have a higher tendency to avoid its own trail due to the negative auto-chemotaxis effect, which results in a scaling of MSD$\sim \Delta t^{1.5}$. It should be noted here that the extent of negative autochemotaxis of self-avoidance strongly depends on the degree of droplet confinement, which we proceed to discuss by considering droplets of varying sizes.

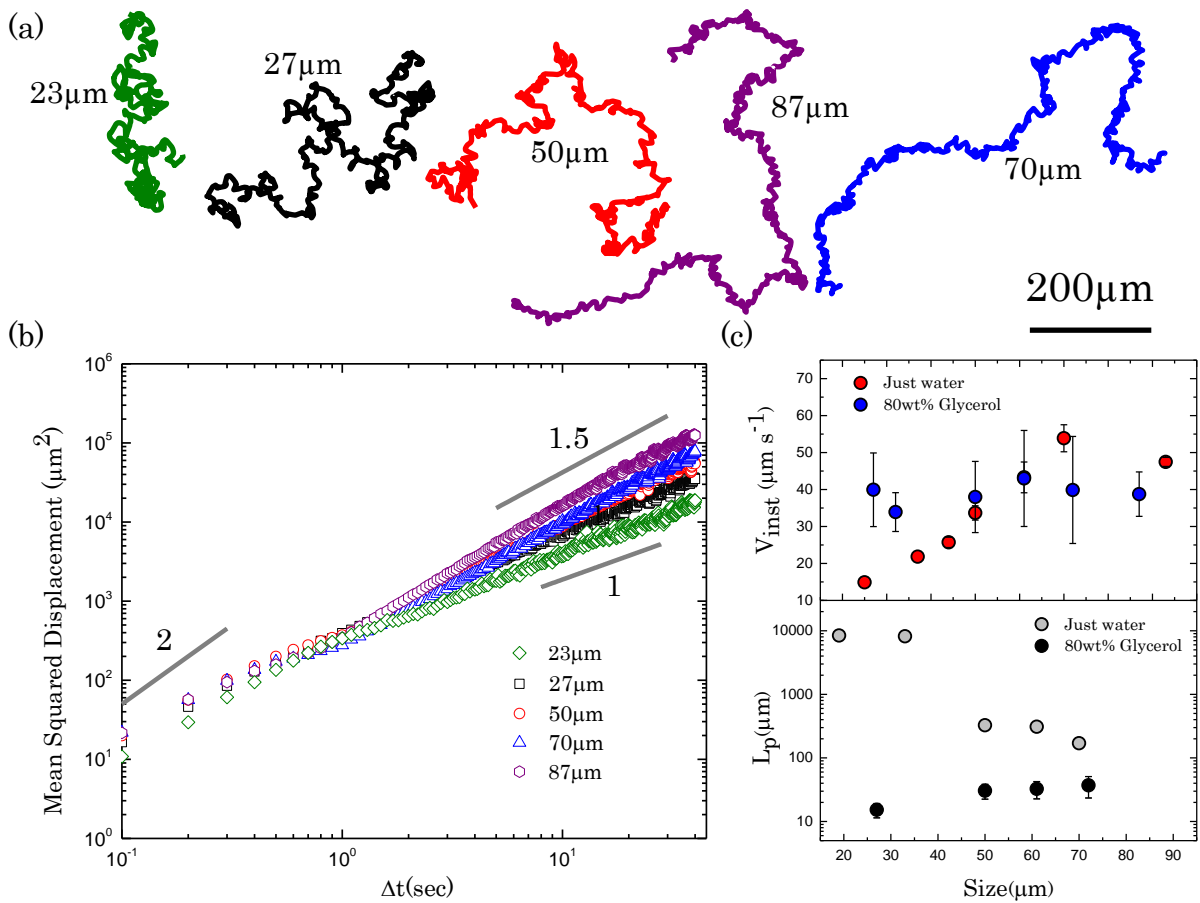

Figure 6: (a) Trajectories ($\Delta t$=100s) of active 5CB droplets in 80wt% glycerol-water TTAB solution with varying size (b) Corresponding mean square displacement with time for trajectories captured over 600s. (c) Instantaneous speed and persistence length of 5CB droplets in aqueous TTAB.

In order to focus on the role of confinement, we carried out experiments of active motion of 5CB droplets for different sizes in 80wt% glycerol-water-TTAB solution. As seen from representative X-Y trajectories shown in figure 6(a), the jittery fluctuations in trajectories were observed for droplets of all different sizes. The corresponding 2D MSD values have been shown in figure 6(b). Unlike active 5CB droplets in just TTAB-water solution (cf. Fig 3b), in 80wt% glycerol-water solution, the temporal increase in MSD ($\sim \Delta t^2$) at short time scales is independent of the droplet size. This observation is further supported by the independent $\langle v_{inst} \rangle$ measurements using the droplets' time dependent position, shown in figure 6(c). Also, considering the variation in the observed values, it appears that $\langle v_{inst} \rangle$ values observed in the presence of glycerol are of the same order as values observed in the absence of glycerol. Since 80wt% glycerol-water solution is nearly 40 times more viscous compared to just water, this is an unexpected observation, which we will discuss later in the article. Despite similar $\langle v_{inst} \rangle$ values, it is clear that the long-time MSD of the 5CB droplet in the presence of glycerol is much lesser than their MSD in the absence of glycerol (see figure 2(b) and figure 6(b)), which suggests that glycerol induces significant fluctuations in the active motion. In addition, similar to the case of water, in presence of glycerol, the slope of MSD at long times shows a size dependent variation. With increasing droplet size, the slope of MSD increases from 1 to 1.5. We speculate the reason for this transition with increasing droplet size to be the difference in the Z-confinement offered by the optical cell to the droplets with different sizes. Supporting movie S4, clearly shows that the smaller droplets occasionally move in Z-direction and go out of focus, whereas, larger droplets remain in focus all the time confirming negligible Z movement. As we are only recording the 2D motion, we are unable to capture the self-avoiding nature of the occasional Z steps taken by the smaller droplets to avoid the trail of filled micelles. Therefore, the long-time 2D MSD fails to demonstrate a scaling of 1.5, which, however, is successfully observed for larger droplets since their motion remains constrained in 2D. This enhanced self-avoiding walk nature of larger droplets is also supported by the variation in $L_p$ as shown in figure 6(c). For comparison, $L_p$ variation for active droplets in just water has also been shown. Consistent with previous observations, addition of glycerol significantly lowers $L_p$ compared to no glycerol case.

Hokmabad *et al.* [47] proposed that addition of glycerol resulted in an increase in viscosity and hence an increase in *Pe,* which induced a new higher order mode of active transport responsible for the fluctuations. To verify the role of *Pe* in this transition to jittery motion, we repeated the experiments with 1wt% PAAm ($M_w$~6000 kDa), 20wt% PVP ($M_w$~40 kDa) and 5wt% PVA ($M_w$~115 kDa) as additives to the aqueous surfactant solution. The viscosity of all the polymeric solutions is similar to the viscosity of 80wt% glycerol aqueous solution, see supporting figure S5. Figures 7(a-c) show a few representative X-Y trajectories in different solutions (see supporting videos S9-S12). With PAAm and PVA as solutes, the anomalous jittery motion of the droplets is absent. In fact, the trajectories mostly resemble the smooth motion observed in solute-free aqueous TTAB solution. Whereas for PVP solution, significant jitteriness in droplet motion is evident. In figure 8, we compare the nature (jittery vs smooth) of the active motion of dispersed droplets under different solute conditions with changing $Pe = R\langle v_{inst}\rangle / D_{surfactant} = 6\pi\eta aR\langle v_{inst}\rangle / k_B T$. In figure 8(a), Pe was experimentally calculated using experimentally observed $\langle v_{inst}\rangle$ values of the active droplets and other relevant physical parameters. Here *a* is the size of a surfactant molecule. In figure 8(b), we used the theoretical formulation of *Pe* discussed by Hokmabad *et al.* (see supporting information). In both figures, circular markers represent systems with regular smooth trajectory, whereas, the star markers represent systems with the unusual jittery trajectory. From these figures, it is clear that despite similar range of *Pe* values observed for 80wt% glycerol and other polymeric solutes, only glycerol and PVP induce anomalous jittery motion in active 5CB droplets, whereas PAAm and PVA do not. This confirms that a mere increase in viscosity, or *Pe*, is not the sole underlying reason behind the fluctuations. Our observations suggest that the line of the argument indicating the transition mechanism to be solely based on Peclet number, as proposed by Morozov and Michein,[55] is perhaps simplistic. In figure 8(b), we have also included the study by Izri *et al.*[17] which reported the active motion of water droplets in an isotropic oil, squalene, using monoolein as a surfactant and did not exhibit any jittery motion. To further understand the underlying mechanism for the observed jittery motion, we focus on the continuous dissolution of the droplet (-dR/dt > 0). Figure 7(d) compares the rate of solubilization of the droplets in 6wt% TTAB aqueous solution with different solutes. It is clear that the occurrence of jittery motion in presence of glycerol and PVP coincides with a higher (~ 3 times) -dR/dt in comparison to the case of no solute or PAAm or PVA as a solute, where motion remains smooth. This suggests that a mechanism based on faster solubilization of the active droplets is responsible for the jittery motion. Since propulsion is directly related to the solubilization mechanism, a higher (-dR/dt) in 80wt% glycerol is consistent with the higher than expected $v_{inst}$ values, despite a significant increase in the viscosity.

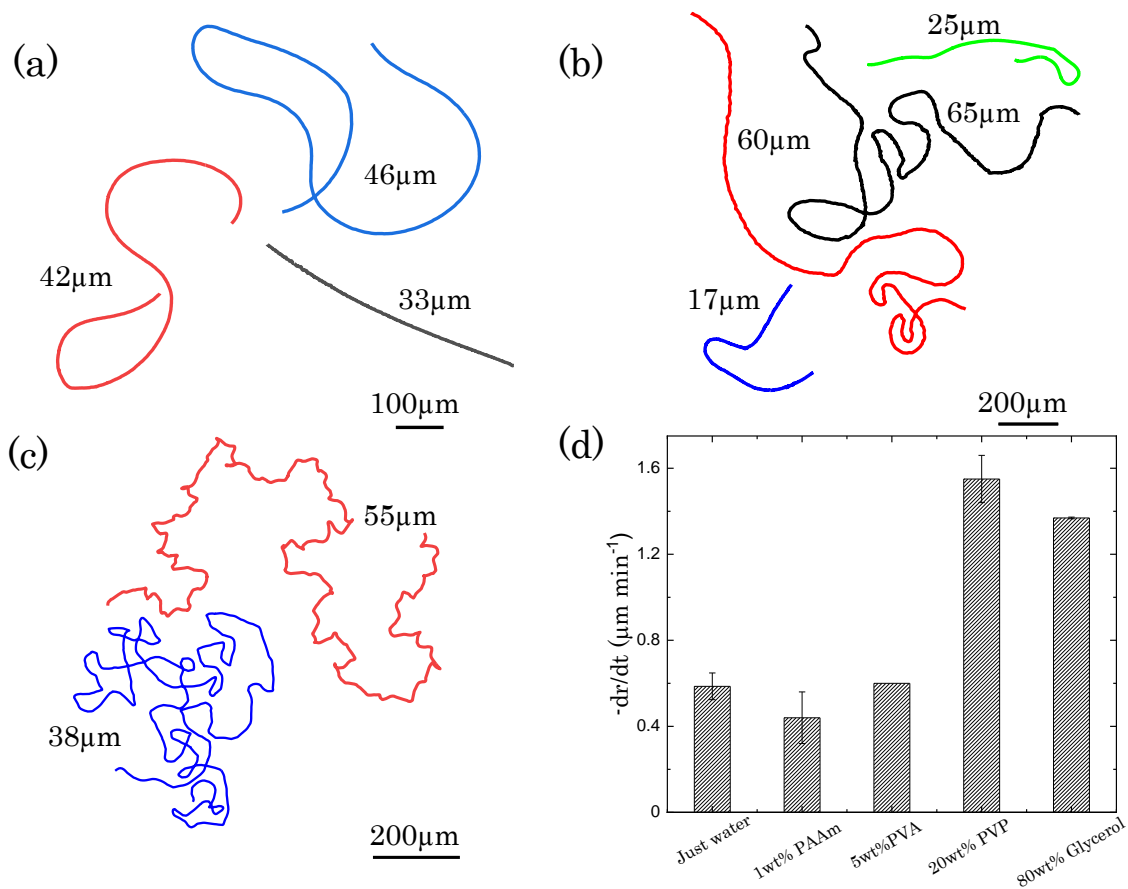

Figure 7: Trajectories active 5CB droplets captured over 100 s (a) 1wt% PAAm aqueous TTAB solution. (b) 5wt% PVA aqueous TTAB solution (c) 20wt% PVP aqueous TTAB solution (d) Rate of solubilization of 5CB droplets in aqueous TTAB solution with presence of various solutes.

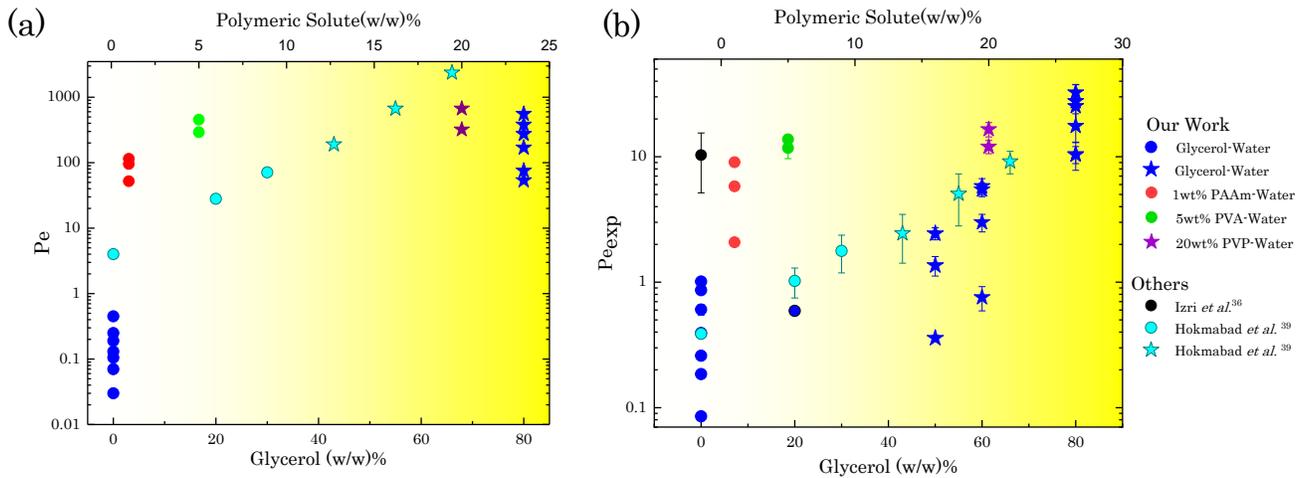

Figure 8: Variation in modes of trajectory for active droplets in different systems with varying Pé number.(a) Pé number calculated using theoretical expression (see supporting information) (b) Pé number based on self propelled speed obtained experimentally.

Proposed mechanism for jitteriness in active motion

Based on the foregoing observations, we now propose a mechanism for the observed jittery droplet motion. The mechanism is illustrated in the schematic shown in figure 9. The mode of self-propulsion of an active droplet depends on two critical time scales, namely, the mean time-scale of surfactant in-flux at the site of adsorption ($\tau_{in-flux}$) and the mean time scale of their migration away ($\tau_{out-flux}$) from the adsorption region. If $\tau_{in-flux} \leq \tau_{out-flux}$, the migration of surfactant away from the leading edge, where fresh surfactants are being adsorbed, is slow. Hence, a surface tension gradient will be maintained, forcing the droplet to perform sustained active motion in the same direction. On the other hand, if $\tau_{in-flux} > \tau_{out-flux}$, the surfactant molecules adsorbed at the leading edge migrate quickly, before the adsorption of another incoming micelle. As a result, the droplet will momentarily experience a loss of surface tension gradient and will come to a halt. The motion is then reinitiated due to a *re*-generation of surface tension gradient caused by the collision with another empty micelle. Since, the collison can happen from an independent direction, other than the earlier wake region, the new direction of motion will result in a pseudo random jittery motion. Using an analytical approach, Schmitt and Stark[41] theoretically predicted the possibility of a random trajectory under similar conditions. To validate the mechanism, we identify the relevant time scales, i.e., $\tau_{in-flux}$ and $\tau_{out-flux}$, in our experiments. We assume that adsorption of surfactant molecules due to the disintegration of free micelles in the vicinity of the droplet is instantaneous. Therefore, the overall rate of surfactant in-flux is limited by advection of the micelles to the apex of the droplet, hence $\tau_{in-flux} \sim \dfrac{R}{\langle v_{inst} \rangle}$. These incident surfactant molecules will migrate away from the apex towards the rear end of the droplet, due to the solubilization (release of swollen micelles) generated Marangoni convection and diffusiophoretic slip at the interface. To measure this migration time scale, we injected small 2 μm (<<R) polystyrene (PS) particles as tracers to measure the interfacial speed. The tracer particles in the vicinity of the actively moving droplet were carried by the resultant fluid flow in the bulk. It was observed that particles very close to the interface experienced a slingshot along the curvature opposite to the direction of motion of the droplet, see figure 10 (a). Using the trajectory of these particles, we estimated the interfacial speed, $v_{int.}$, at the interface using which the $\tau_{out-flux} \sim \frac{R}{v_{int.}}$ was calculated. As shown in figure 10(b), the ratio of these time scales, $\tau_{out-flux}/\tau_{in-flux}$, demonstrates a monotonic decrease with glycerol concentration, in support of our hypothesis. In addition, in figure 10(c), we show the cross-polarized time lapse images of an active 5CB droplet in 80wt% glycerol. The images confirm the presence of a radial hedgehog defect in the 5CB droplet due to TTAB surfactant layer generating a homeotropic anchoring of the nematic director. Images can be classified in two categories: (a) where the defect is symmetrically located at the center, which is consistent with the resting state of the droplet (see figure 10(c) and (d)), and (b) where the defect appears to de dislocated towards the edge, which is a characteristic of a self-propelling droplet. The position of the discloated defect points to the direction of motion.[46] By extension, change in the location of the dislocated defects, before and after the resting state, implies change in direction of the motion of the droplet, which is consistent with their fluctuation motion. These observations further corroborate our proposed mechanism.

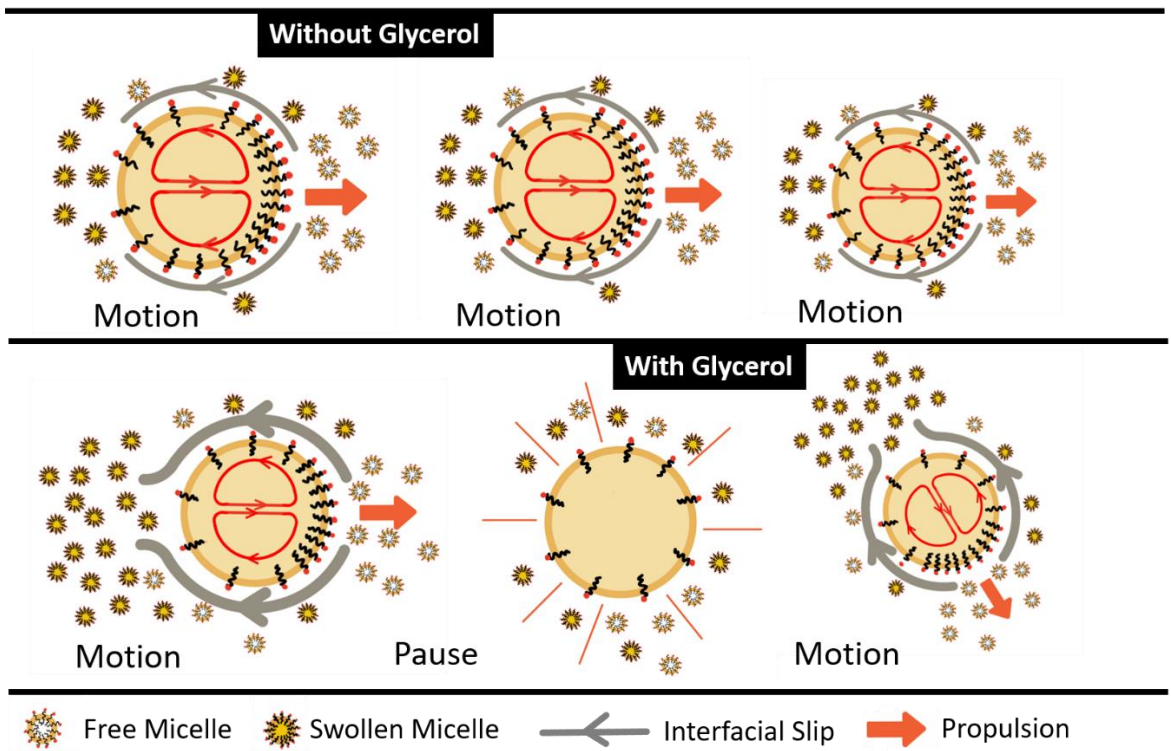

Figure 9: Schematic representation of the proposed mechanism behind the jittery motion observed in glycerol-water solution, in comparison to the usual unperturbed motion in absence of glycerol.

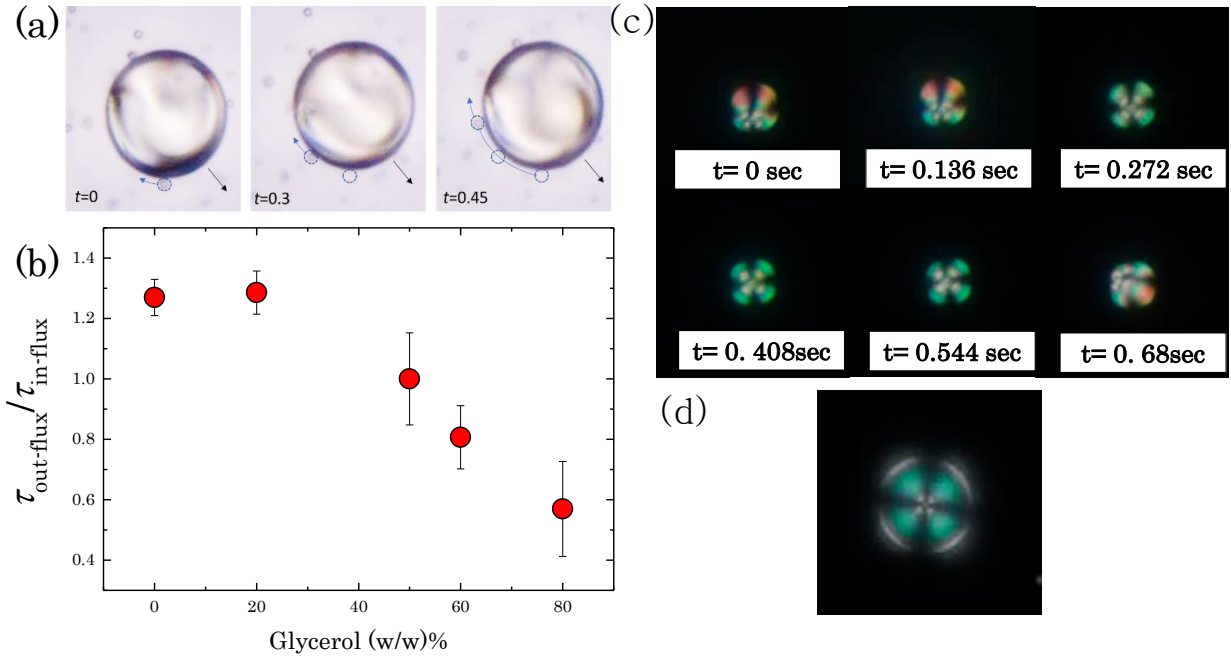

Figure 10: (a) Time series snapshots of a 5CB active droplet (~60 μm) in 6wt% TTAB aqueous solution with 80wt% glycerol, with a tracer particle (2 μm PS) near the interface (see supporting video S13). (b) Ratio of $\tau_{out\text{-}flux}$ and $\tau_{in\text{-}flux}$ with varying glycerol concentration. (c) Cross-polarized time lapse images of moving 5CB liquid crystal droplet in aqueous glycerol (80wt%) TTAB (6wt%) solution. (b) Cross-polarized image of a stagnant 5CB liquid crystal droplet in aqueous glycerol (80wt%) solution with 1wt% TTAB.

To understand the faster migration of surfactant molecules at the 5CB droplet interface, we focus on the physico-chemical effect resulting from the addition of glycerol. Although the exact surfactant concentration profile at the droplet's interface is unknown, the leading-edge is expected to be at surfactant surface concentration (Γ) close to $Γ_∞$ and the trailing-edge with $Γ << Γ_∞$, where $Γ_∞$ is the surface-excess concentration at the interface in equilibrium to the bulk surfactant concentration of CMC. A rough estimate of Marangoni flow is $v_M \sim \frac{\left|d\gamma/dc\right|}{\eta_i + \eta_o} \nabla_s c \sim \frac{\gamma_o - \gamma_{CMC}}{\eta_i + \eta_o}$. Here, $\gamma_o$ is the interfacial tension in absence of surfactant ($\Gamma << \Gamma_\infty$) and $\gamma_{CMC}$ is the interfacial tension corresponding to surface excess concentration of $\Gamma_\infty$. The interfacial tension ($\gamma$) of 5CB in aqueous solution with varying TTAB concentration ($c_{TTAB}$) shows a linear decrease in $\gamma$ at low $c$ followed by a plateau at higher $c$ for all solutes (see figure S1). The concentration where the transition from a rapid decrease to a plateau region occurs is the CMC of TTAB in the respective continuous media.

Based on these measurements, we expect that $v_{M,water} > v_{M,glycerol}$, which is in contradiction to the experimental observations. We hypothesize, two possible reasons behind this:

  a) The accumulation of swollen micelles close to the droplet interface leads to a reduced concentration of free micelles which has been previously reported to be responsible for negative-auto chemotaxis.[30] Active oil droplets have been shown to avoid their trails filled with swollen micelles, and instead move towards untraced paths. Recent numerical studies[38] have reported that the accumulation of free swollen micelles inhibit the transport of fresh surfactants to the interface. Therefore, we expect this phenomenon to contribute towards inhomogeneous surfactant adsorption at the interface. Since, in the presence of glycerol, the parent droplet releases more swollen micelles compared to that in solute-free aqueous surfactant solution, the former will experience a higher $\nabla c$ between the leading and the trailing edge.

  b) The faster rate of solubilization of 5CB droplet in the presence of excess of glycerol leads to a faster release of swollen micelles. These swollen micelles accumulate near the trailing edge and generate a non-uniform concentration gradient across the droplet. We believe that the gradient of these released solutes (non-attractive) also creates a chemiosmotic slip which further enhances the transport of the surfactant molecules at the interface. Overall, the effective Marangoni stresses are amplified resulting in higher velocity at the interface. For a solvophobic solid wall, it has been earlier shown that diffusio-osmosis can enhance the wall slip by 10-100 times.[56] In our case, as it is a fluid-fluid interface, a similar rise in interfacial transport is not expected. However, the resultant diffusion-osmotic flow is expected to result in an enhanced Marangoni convection at the interface.

Finally, we set out to discuss the possible reasons for the observed higher rate of solubilization in the presence of glycerol. The process of micellar solubilization is governed by intermolecular interactions between different species and thermodynamic processes involved. Therefore, it is well established that the process is extremely sensitive to numerous physical and chemical factors including temperature, pressure, pH and presence of additives. Typically, additives which affect the CMC and the aggregation number in a micelle, noticeably, are expected to produce parallel effects on solubilization as well. In ionic surfactant solutions, it has been observed that the addition of polar additives such as phenols and other alcohols can enhance the micellar solubility. It has been proposed that such additives insert themselves between the adjacent surfactant molecules in a micelle and they act as co-surfactants.[40] As discussed in the earlier sections and observed in our experiments, the addition of glycerol was accompanied by modification of surfactant activity. Therefore, the discussion above is suitable to our experimental observations and it is likely that the addition of glycerol significantly enhances the rate of micellar solubilization by altering the thermodynamic processes involved. This is clearly evident from the change in CMC of TTAB with addition of glycerol in water. The CMC in 80wt% glycerol-water solution, $CMC_{GW} \sim 0.45wt\%$, which is significantly hihger than the CMC in pure water (CMC$_{water}$ ~0.13%). A higher CMC in the presence of glycerol hints towards higher solubility of TTAB, consistent with previous study by Moya *et al.*.[51] The study also reported that a higher solubility of TTAB in glycerol resulted in an increased aggregation number ($N_{agg}$) of TTAB, i.e., number of TTAB surfactant molecules present in a single micelle. In water, $N_{agg}$ is ~30, which increases to ~70 for the case of ~80wt% glycerol-water solution. Additionally, we observed that the interfacial tension of 5CB in aqueous solution of TTAB is always lower in the presence of glycerol compared to pure water. In fact, glycerol has been used as a co-solvent to enhance oil solubilization in spontaneous emulsification.[57] Thus, we conclude that the observed jittery motion of active 5CB droplets in the presence of glycerol is primarily a consequence of glycerol enhanced micellar solubilization. In the presence of PVP as a solute, similar physicochemical effects are expected to lead to an enhanced rate of solubilization for the 5CB droplets in TTAB aqueous solution, leading to jittery motion.

## CONCLUSIONS

In summary, we have investigated the effect of addition of different solutes on the active transport of 5CB liquid crystal (oil) droplets in aqueous surfactant (6wt% TTAB) solution. Unlike the smooth active motion observed in TTAB aqueous solution, addition of glycerol (≥50wt%) as solute, resulted in a peculiar jittery motion, where the fluctuations in the trajectories are observed for all different droplet sizes investigated. Additional experiments using equi-viscous TTAB aqueous solution with PAAm, PVP and PVA as solutes were also performed. Interestingly, while the presence of glycerol and PVP as solutes induce a transition to jittery motion, an equi-viscous solution of PAAm and PVA fail to do so. Another key observation is that, in the presence of glycerol and PVP, the rate of solubilization of the active droplets is noticeably higher in comparison to that observed in solute-free aqueous solution or in the presence of PAAm/PVA as solutes. While Hokmabad *et al.* suggest that an increase in viscosity and thereby Peclet number causes a transition from quasi-ballistic propulsion to jittery motion, our experiments with additional polymeric solutes PAAm, PVA and PVP clearly indicate that a more complex mechanism related to the rate of solubilization is at the heart of this transition. Our equal Peclet number experiments with glycerol and polymeric solutes also indicate toward this conclusion.

Our experiments suggest that the nature of physicochemical interaction of the solute with that of the surfactant and its eventual effect on droplet solubilization determines whether such a transition can occur. By altering the activity of the surfactant molecules, evident from the increase in CMC and reduction in saturation interfacial tension for 5CB-surfactant solution interface, addition of glycerol, enhances micellar solubilization. By exploring the tracer particles in the vicinity of the active droplets, we confirmed that the enhanced micellar solubilization results in faster migration of the surfactant molecules at the interface in comparison to their influx time scale. Under these circumstances, the surfactant concentration gradient quickly re-homogenizes, bringing the droplet to a halt, until the motion is re-initiated from an independent direction. This hypothesis was further supported by the captured time series of cross-polarized images of the active 5CB droplets in presence of glycerol. More experiments and detailed theoretical efforts can further provide insights into the origin on the nature of this transition.

## AUTHOR INFORMATION

Corresponding Author

* Email: mangalr@iitk.ac.in

Funding Sources

This work is supported by the Science and Engineering Research Board (SB/S2/RJN-105/2017), Department of Science and Technology, India.

ACKNOWLEDGMENT

We thank Prof. Jayant K. Singh for interfacial tension measurements. We also thank Prof. Naveen Tiwari and Prof. Harshwardhan Katkar for useful discussions.